\documentclass[letterpaper,12pt]{article}

\usepackage{ol2_edit2}
\usepackage{enumitem}
\usepackage{graphicx}
\usepackage{psfrag}
\usepackage{subfigure}
\usepackage{floatflt}
\usepackage{wrapfig}

\begin{document}


\title{MEMS-Based Optical Beam Steering System for Quantum Information Processing in 2D Atomic Systems}

\author{Caleb Knoernschild,$^{1,*}$ Changsoon Kim,$^1$ Bin Liu,$^1$ Felix P. Lu,$^{1,2}$ Jungsang Kim$^1$}
\address{
$^1$Fitzpatrick Institute for Photonics, Electrical and Computer Engineering Department, Duke University, Durham NC, 27708\\
$^2$Applied Quantum Technologies, Inc., Durham NC, 27708\\
$^*$Corresponding author: caleb.k@duke.edu
}


\begin{abstract}
\textit{In order to provide scalability to quantum information processors utilizing trapped atoms or ions as quantum bits (qubits), the capability to address multiple individual qubits in a large array is needed.  Micro-electromechanical systems (MEMS) technology can be used to create a flexible and scalable optical system to direct the necessary laser beams to multiple qubit locations.  We developed beam steering optics using controllable MEMS mirrors that enable one laser beam to address multiple qubit locations in a 2 dimensional trap lattice.  MEMS mirror settling times of {$\sim10\mu$}s were demonstrated which allow for fast access time between qubits.}  
\end{abstract}

\maketitle


\noindent Some of the most promising physical implementations of quantum information processors (QIPs) utilize internal states of ions coupled via Coulomb interactions\cite{CiracPRL1995,MonroePRL1995,Schmidt-KalerN2003,LeibfriedN2003} or neutral atoms coupled through dipole-dipole interactions\cite{JakschPRL2000,YavuzPRL2006} to represent quantum bits (qubits). Manipulation of qubit states in these QIP implementations require precisely controlled laser beams. While architectures for scalable QIPs have been proposed,\cite{KielpinskiN2002} their realization is limited by available technology.\cite{KimQI&C2005} To improve the scalability of these experiments, an effective means of delivering laser beams to multiple qubit locations is required. Requirements such as fast addressing times ({$\sim1\mu$}s) imposed by qubit dephasing,\cite{SaffmanPRA2005} broad range of operational wavelengths (UV for trapped ions\cite{MonroePRL1995,Schmidt-KalerN2003,LeibfriedN2003} and IR for neutral atoms\cite{SaffmanPRA2005,PortoRS2003}), and scalable addressing of thousands of locations exclude the possibility of using traditional optical components. Acousto-optical modulators have been used to provide beam steering to several qubit locations\cite{Schmidt-KalerAPB2003}, but this approach introduces a frequency shift and is difficult and costly to scale.  Optical micro-electromechanical systems (MEMS) can provide a variety of optical functionality such as beam steering \cite{NeilsonJoLT2004,HicksOL2007} and focus control \cite{LiuOL2003} in a highly compact and integrable form, and have been suggested as viable solutions to meet the needs of atomic based QIPs.\cite{KimQI&C2005,Metodi2005} We report a scalable MEMS-based 2 dimensional (2D) beam steering system capable of addressing a 5$\times$5 array of trapped atoms that can be easily modified to address trapped ions. In our system, diffraction limited optical design minimizes optical power at neighboring lattice locations while analog control electronics provide full system reconfigurability for random access of the quantum bits (qubits) in the lattice.

\begin{figure}[htb]
\centering
\includegraphics[width=8.25cm]{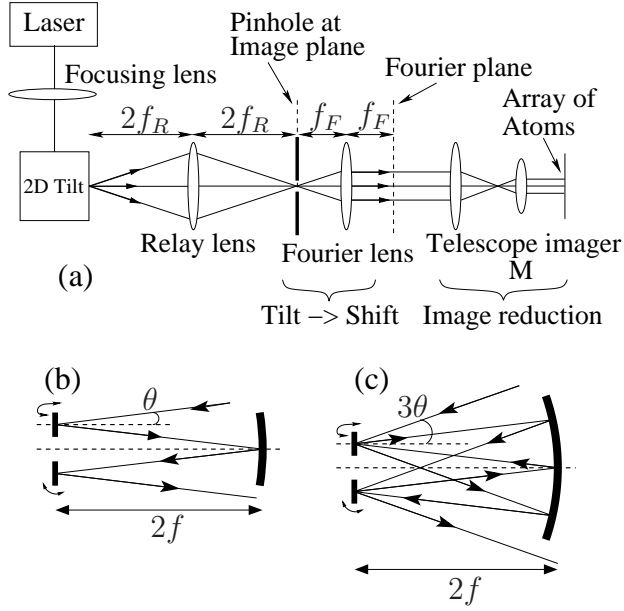}
\caption{\label{fig:sys} (a) Schematic of the beam steering system (b) 2$f$-2$f$ folding imaging optics to combine decoupled tilt motion for a
single bounce system  (c) Double bounce system.}
\end{figure}

Fig.~\ref{fig:sys}a shows the schematic of the beam steering system. Current neutral atom experiments use optical lattices with lattice constant of {$a \approx 8 \mu$}m.\cite{YavuzPRL2006} In order to provide access to each atom in the array with minimal intensity at neighboring locations, we choose the beam waist at the atom to be {$w_o = a/2$}. To manage the laser divergence in the beam steering system, a magnifying telescope imager is used to transform the lattice constant by a factor of {$M$} to {$a'=Ma$}. and the corresponding beam waist to {$w_o'=Mw_o=Ma/2$} at the ``Fourier plane'' defined in Fig.~\ref{fig:sys}a. Tilting MEMS mirrors are utilized in our system to provide beam steering to the 25 lattice locations of the 5$\times$5 array. The beam reflecting off tilted mirrors is imaged through a {$2f_R$}-{$2f_R$} relay lens onto an image plane to allow for easier optical alignment. Stray reflections from the MEMS package window are filtered out using a pinhole located at the image plane. The angular tilt of the beam is converted into parallel lateral shifts by a Fourier lens located a focal length {$f_F$} away from the image plane, and then projected onto the target array through the telescope imager. 

The 2D tilt motion is decoupled into two separate 1D MEMS mirrors, tilting in orthogonal directions, by using a spherical mirror in a folded {$2f$}-{$2f$} imaging system to refocus the reflected beam from the first mirror to the second (Fig.~\ref{fig:sys}b). This optical arrangement enables an angle multiplication scheme where the maximum output angular range is magnified for a given range of mirror tilt angles. Increasing the incident angle {$\theta$} at the first mirror to {$(2n-1)\theta$} induces $n$ bounces per MEMS mirror. The mechanical tilt angle of the mirror, $a$, is translated to the beam tilt angle of {$2na$} at the output. We implement a double bounce system ({$n=2$}) shown in Fig.~\ref{fig:sys}c to double the output angular range of a single bounce system. Larger multiplication increases spherical aberration and reduces throughput of the system when mirror reflectivity is below unity. The entire optical system was modeled using Zemax optical simulation software to accurately assess performance.

\begin{figure}[htb]
\centering
\includegraphics[width=6.5cm]{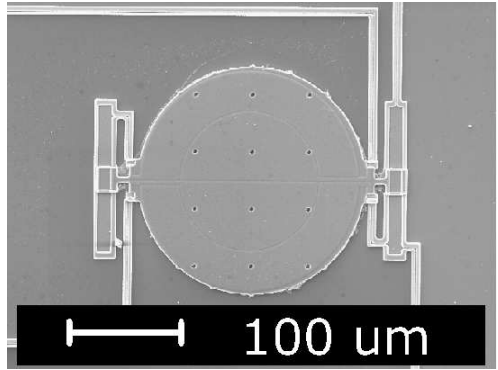}
\caption{\label{fig:sem} Scanning electron micrograph of MEMS mirror with radius of {$100 \mu$}m.}
\end{figure}

The MEMS mirror consists of a mirror plate that rotates about two torsional springs (Fig.~\ref{fig:sem}) and is actuated electrostatically by means of a grounded mirror plate and underlying electrodes. The fabrication was done using the PolyMUMPS foundry process at MEMScAP, Inc.\cite{Memscap} The mirror was coated with aluminum to enhance reflectivity and placed in a package sealed with a quartz window. The dynamic characteristics of the MEMS mirrors can be modeled as a damped harmonic oscillator with resonance frequency {$\omega=\sqrt{2K/I}$} where {$K$} is the torsional stiffness for one of the mirror's springs and {$I$} is the moment of inertia of the mirror plate. Because neutral atom gate operations require {$\sim1\mu$}s switching times, we designed our mirrors to minimize settling time by increasing {$\omega$} and maintaining near critical damping.

Mirror size and maximum tilt angle are determined by the optical system.  Defining a variable {$\zeta=f_F/M$} and incorporating an {$n$} bounce system, the beam waist at the MEMS mirror ({$w_M$}) and the maximum required mechanical tilt {$(\Delta\theta)_{max}$} are given by \begin{equation}
 w_M=\frac{\lambda f_F}{\pi w_o'}=\frac{\lambda}{\pi w_o}\zeta
\end{equation}
\begin{equation}
 (\Delta\theta)_{max}=\frac{a'}{f_F}\frac{N-1}{4n}=\frac{a}{\zeta}\frac{N-1}{4n}
\end{equation}
where {$\lambda=780$}nm is the target wavelength and {$N=5$} is the size of the lattice in 1D. Solving 1 and 2 for {$\zeta$}, we see that {$w_M\propto 1/(\Delta\theta)_{max}$}. Physical constraints on actuation voltages, realizable spring stiffness, and the mirror size to tilt angle relationship restrict the design space for settling time optimization.\cite{ChangsoonSTiQEIJo2007} The angle multiplication scheme relaxes the design space by reducing {$(\Delta\theta)_{max}$} while maintaining the same mirror size.  We could further reduce {$(\Delta\theta)_{max}$} by increasing {$n$}, but this would require higher mirror reflectivity to compensate for the reflection losses in the multiple bounce system.  

\begin{figure}[htb]
\centering
\includegraphics[width=7.5cm]{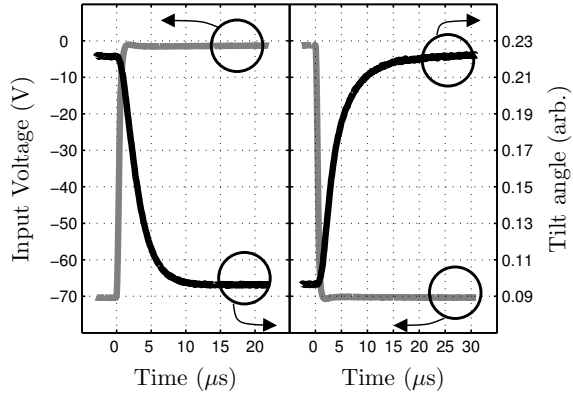}
\caption{\label{fig:trans} Transient responses featuring settling time of {$15 \mu$}s (right) and {$10 \mu$}s (left). A shift from the central position to an adjacent location (right) and back (left) requires {$90$}V for full angular range in one direction.  }
\end{figure}

Mirrors with radii of {$60$}, {$75$}, and {$100 \mu$}m were fabricated with various {$K$} values. The transient response of a typical {$100 \mu$}m radius system mirror is shown in Fig.~\ref{fig:trans} along with the driving voltage. Mirrors with settling times of {$\sim 10 \mu$}s were used in the beam steering system (Fig.~\ref{fig:trans}). The settling time when an actuation voltage is used to drive the mirror is longer than when an applied voltage is removed\cite{ChangsoonSTiQEIJo2007} due to electrostatic effects, which can be compensated for by slight tailoring of the driving voltage shape.

\begin{figure}[htb]
\centering
\includegraphics[width=7.5cm]{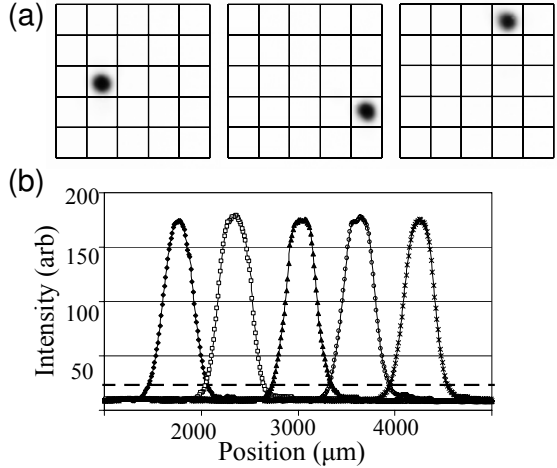}
\caption{\label{fig:spots} (a) Three spots within 5$\times$5 addressing array. (b) Intensity profile of 5 neighboring positions.  Dashed line displays the {$1/e^{2}$} point.}
\end{figure}

Intensity profile measurements at a wavelength of {$780$}nm were taken at the Fourier lens focal plane (Fourier plane) prior to spot size demagnification. Fig.~\ref{fig:spots} shows three shifted laser locations within the 5$\times$5 array (a) and the shifted intensity profiles of five neighboring locations (b). All 25 positions were addressed with stable intensity profiles. Residual intensities at neighboring qubit locations were measured to be more than {$30$}dB below the peak output intensity, consistent with Gaussian beam predictions. The shifted laser profiles demonstrate a complete beam diameter shift to neighboring locations and peak to peak distances of twice the output beam waist. Analog actuation voltages driven by digital circuits allow for easy controllability and fine shifting control. Beam distortion of the Gaussian mode shape due to the system was not noticeable. Introducing beam reduction optics can easily scale the measured results at the Fourier plane down to the necessary dimensions at the atom locations.

Using a plano-convex meniscus lens combination for the relay lens to reduce spherical aberration, we measured peak intensity variations of {$<5\%$}. System simulation in Zemax shows intensity variation of  {$\sim3\%$} indicating small room for improvement. We observed residual unwanted reflections from the package window which can be eliminated through better anti-reflection coating. The variation of peak intensities at this level does not introduce any substanctial increase in the beam radius at the atom locations which would impact neighboring atoms. The effect of these variations on qubit manipulation can be compensated by adjusting the duration of the beam interaction with the atoms for each site.  
  
The optical throughput of the system suffers from losses in reflectivity from the MEMS mirror and stray reflections from the package window. The aluminum coating used on the mirror has a dip in reflectivity around {$780$}nm ({$<80\%$}). We measured system throughputs of {$43\%$} for a single bounce system and {$24\%$} for a double bounce system.  Mirror reflectivity can be enhanced to {$>90\%$} with optimized coating, which will allow {$>50\%$} system throughput for the double bounce system.

While mirrors with settling time of {$\sim 10 \mu$}s were used for system demonstration, we have fabricated and tested mirrors with settling times {$<3 \mu$}s. Our modeling indicates that further reduction of settling times ({$1$}-{$2 \mu$}s) is possible by device optimization.  

In summary we developed a MEMS based optical beam steering system for a QIP utilizing neutral atoms or trapped ions as qubits. A 25 spot lattice was individually addressed with minimal optical intensity at neighboring positions with mirror settling times {$\sim10 \mu $}s. This technology is versatile and can be adjusted to support multiple laser colors and varying output array patterns. 

This work was supported by NSF under CCF-0520702 and ARO STTR Program under W911NF-06-C-0112. The authors would like to thank John Foreman and Henry Everitt for their help in measuring reflectivity of MEMS mirrors in the UV wavelength range.


\begin{thebibliography}{10}
\newcommand{\enquote}[1]{``#1''}

\bibitem{CiracPRL1995}
J.~I. Cirac and P.~Zoller, \enquote{Quantum computations with cold trapped
  ions,} Phys. Rev. Lett. \textbf{74}, 4091 (1995).

\bibitem{MonroePRL1995}
C.~Monroe, D.~M. Meekhof, B.~E. King, W.~M. Itano, and D.~J. Wineland,
  \enquote{Demonstration of a fundamental quantum logic gate,} Phys. Rev. Lett.
  \textbf{75}, 4714--4717 (1995).

\bibitem{Schmidt-KalerN2003}
F.~Schmidt-Kaler, H.~Haffner, M.~Riebe, S.~Gulde, G.~P.~T. Lancaster,
  T.~Deuschle, C.~Becher, C.~F. Roos, J.~Eschner, and R.~Blatt,
  \enquote{Realization of the cirac-zoller controlled-not quantum gate,} Nature
  \textbf{422}, 408--411 (2003).

\bibitem{LeibfriedN2003}
D.~Leibfried, B.~DeMarco, V.~Meyer, D.~Lucas, M.~Barrett, J.~Britton, W.~M.
  Itano, B.~Jelenkovic, C.~Langer, T.~Rosenband, and D.~J. Wineland,
  \enquote{Experimental demonstration of a robust, high-fidelity geometric two
  ion-qubit phase gate,} Nature \textbf{422}, 412--415 (2003).

\bibitem{JakschPRL2000}
D.~Jaksch, J.~I. Cirac, P.~Zoller, S.~L. Rolston, R.~Cote, and M.~D. Lukin,
  \enquote{Fast quantum gates for neutral atoms,} Phys. Rev. Lett. \textbf{85},
  2208--2211 (2000).

\bibitem{YavuzPRL2006}
D.~D. Yavuz, P.~B. Kulatunga, E.~Urban, T.~A. Johnson, N.~Proite, T.~Henage,
  T.~G. Walker, and M.~Saffman, \enquote{Fast ground state manipulation of
  neutral atoms in microscopic optical traps,} Phys. Rev. Lett. \textbf{96}
  (2006).

\bibitem{KielpinskiN2002}
D.~Kielpinski, C.~Monroe, and D.~J. Wineland, \enquote{Architecture for a
  large-scale ion-trap quantum computer,} Nature \textbf{417}, 709--711 (2002).

\bibitem{KimQI&C2005}
J.~Kim, S.~Pau, Z.~Ma, H.~R. McLellan, J.~V. Gates, A.~Kornblit, R.~E. Slusher,
  R.~M. Jopson, I.~Kang, and M.~Dinu, \enquote{System design for large-scale
  ion trap quantum information processor,} Quantum Information \& Computation
  \textbf{5}, 515--537 (2005).

\bibitem{SaffmanPRA2005}
M.~Saffman and T.~G. Walker, \enquote{Analysis of a quantum logic device based
  on dipole-dipole interactions of optically trapped rydberg atoms,} Physical
  Review A \textbf{72} (2005).

\bibitem{PortoRS2003}
J.~V. Porto, S.~Rolston, B.~L. Tolra, C.~J. Williams, and W.~D. Phillips,
  \enquote{Quantum information with neutral atoms as qubits,} Philos. Trans. R.
  Soc. London, Ser. A \textbf{361}, 1417--1427 (2003).

\bibitem{Schmidt-KalerAPB2003}
F.~Schmidt-Kaler, H.~Haffner, S.~Gulde, M.~Riebe, G.~P.~T. Lancaster,
  T.~Deuschle, C.~Becher, W.~Hansel, J.~Eschner, C.~F. Roos, and R.~Blatt,
  \enquote{How to realize a universal quantum gate with trapped ions,} Applied
  Physics B-Lasers and Optics \textbf{77}, 789--796 (2003).

\bibitem{NeilsonJoLT2004}
D.~T. Neilson, R.~Frahm, P.~Kolodner, C.~A. Bolle, R.~Ryf, J.~Kim, A.~R.
  Papazian, C.~J. Nuzman, A.~Gasparyan, N.~R. Basavanhally, V.~A. Aksyuk, and
  J.~V. Gates, \enquote{256 x 256 port optical cross-connect subsystem,}
  Journal of Lightwave Technology \textbf{22}, 1499--1509 (2004).

\bibitem{HicksOL2007}
R.~A. Hicks, V.~T. Nasis, and T.~P. Kurzweg, \enquote{Programmable imaging with
  two-axis micromirrors,} Optics Letters \textbf{32}, 1066--1068 (2007).

\bibitem{LiuOL2003}
W.~Liu and J.~J. Talghader, \enquote{Current-controlled curvature of coated
  micromirrors,} Optics Letters \textbf{28}, 932--934 (2003).

\bibitem{Metodi2005}
T.~S. Metodi, D.~D. Thaker, A.~W. Cross, F.~T. Chong, and I.~L. Chuang,
  \enquote{A quantum logic array microarchitecture: scalable quantum data
  movement and computation,} in \enquote{Proceedings of the 38th annual
  IEEE/ACM International Symposium on Microarchitecture,}  (Barcelona, Spain,
  2005), p. 12 pp.

\bibitem{Memscap}
MEMScAP, \enquote{http://www.memscap.com,} .

\bibitem{ChangsoonSTiQEIJo2007}
C.~Kim, C.~Knoernschild, B.~Liu, and J.~Kim, \enquote{Design and
  characterization of mems micromirrors for ion-trap quantum computation,} IEEE
  J. Sel. Top. Quantum Electron. \textbf{13}, 322--329 (2007).
\end{thebibliography}
\end{document}